\documentclass[universe,article,accept,moreauthors,pdftex]{Definitions/mdpi} 

\graphicspath{{Definitions/Fig/}}
\firstpage{1} 
\pubvolume{1}
\issuenum{1}
\articlenumber{0}
\pubyear{2021}
\copyrightyear{2021}
\externaleditor{Academic Editor: Lorenzo Iorio} 
\datereceived{22 June 2021}
\dateaccepted{21 July 2021} 
\datepublished{} 
\hreflink{https://doi.org/} % If needed use \linebreak
\Title{Are GRMHD Mean-Field Dynamo Models of Thick Accretion Disks SANE?}

\TitleCitation{Are GRMHD Mean-Field Dynamo Models of Thick Accretion Disks SANE?}

\Author{Niccolò Tomei $^{1,2,3,}$*\orcidA{}, Luca Del Zanna $^{1,2,3}$\orcidB{}, Matteo Bugli $^{4}$\orcidC{} and Niccolò Bucciantini $^{2,1,3}$\orcidD{}}

\AuthorNames{Niccolò Tomei, Luca Del Zanna, Matteo Bugli, Niccolò Bucciantini}%please notice that 'Del Zanna' is a single surname, Del is not a prefix or a second name
\AuthorCitation{Tomei, N.; Del Zanna, L.; Bugli, M.; Bucciantini, N.}
\address{%
$^{1}$ \quad Dipartimento di Fisica e Astronomia, Università degli Studi di Firenze, Via G. Sansone 1, \mbox{I-50019 Sesto Fiorentino, Italy}; niccolo.tomei@unifi.it, luca.delzanna@unifi.it\\%emails confirmed, I placed them where the primary affiliation is pertinent (for Bucciantini the main one is number 2)
$^{2}$ \quad INAF, Osservatorio Astrofisico di Arcetri, Largo E. Fermi 5, I-50125 Firenze, Italy ;  	niccolo.bucciantini@inaf.it\\ %Firenze or Sesto Fiorentino are cities, province is not necessary, I removed 'Firenze' as a province where not necessary
$^{3}$ \quad INFN, Sezione di Firenze, Via G. Sansone 1, I-50019 Sesto Fiorentino, Italy \\
$^{4}$\quad ~Département d’Astrophysique, CEA---Saclay, Orme des Merisiers, F-91191 Gif-sur-Yvette, France; matteo.bugli@cea.fr}

\corres{Correspondence: niccolo.tomei@unifi.it}

\abstract{The remarkable results by the Event Horizon Telescope collaboration concerning the emission from M87* and, more recently, its polarization properties, require an increasingly accurate modeling of the plasma flows around the accreting black hole. Radiatively inefficient sources such as M87* and Sgr A* are typically modeled with the SANE (standard and normal evolution) paradigm, if the accretion dynamics is smooth, or with the MAD (magnetically arrested disk) paradigm, if the black hole's magnetosphere reacts by halting the accretion sporadically, resulting in a highly dynamical process. While the recent polarization studies seem to favor MAD models, this may not be true for all sources, and SANE accretion surely still deserves attention. In this work, we investigate the possibility of reaching the typical degree of magnetization and other accretion properties expected for SANE disks by resorting to the mean-field dynamo process in axisymmetric GRMHD simulations, which are supposed to mimic the amplifying action of an unresolved magnetorotational instability-driven turbulence. We show that it is possible to reproduce the main diagnostics present in the literature by starting from very unfavorable initial configurations, such as a purely toroidal magnetic field with negligible magnetization.}


\begin{document}
%%%%%%%%%%%%%%%%%%%%%%%%%%%%%%%%%%%%%%%%%%
%The order of the section titles is: Introduction, Materials and Methods, Results, Discussion, Conclusions for these journals: aerospace,algorithms,antibodies,antioxidants,atmosphere,axioms,biomedicines,carbon,crystals,designs,diagnostics,environments,fermentation,fluids,forests,fractalfract,informatics,information,inventions,jfmk,jrfm,lubricants,neonatalscreening,neuroglia,particles,pharmaceutics,polymers,processes,technologies,viruses,vision

\section{Introduction}

A current challenge in astrophysics is the accurate modeling of the accretion of magnetized plasma onto supermassive black holes. The~surprising observational results by the EHT (Event Horizon Telescope) project have opened the possibility to directly test  sophisticated theoretical models, typically general relativistic magnetohydrodynamical (GRMHD) numerical simulations, against~observations. The~asymmetric ring image of the M87 galaxy's compact radio source was reproduced by ray tracing codes in the curved spacetime of the black hole region, using the emission properties of the extremely hot plasma arising from GRMHD simulations \citep{collaboration_first_2019}. The~further processing of polarized emission on the event horizon scale has allowed an even more accurate estimation of the fundamental physical properties of the accreting plasma, such as density, magnetic field intensity, and electron temperature \citep{akiyama_first_2021}. 

The study of the polarization properties of M87* radio emission has also allowed for considerable progress in the choice of the correct modeling paradigm for the accretion dynamics. It is known that for RIAF (radiatively inefficient accretion flow \citep{narayan_advection-dominated_1994, blandford_fate_1999, narayan_self-similar_2000, quataert_convection-dominated_2000}) sources such as M87* and even SgrA*, the~horizon-penetrating (normalized) magnetic flux $\phi$ can at most increase up to a maximum threshold $\phi_{\text{max}}$, which depends on the spin of the black hole and the scale height of the disk \citep{tchekhovskoy_general_2012}. If~$\phi\sim\phi_{\text{max}}$, the~black hole is unable to continuously absorb  the whole magnetic field, since the accumulation of such a field in its magnetosphere significantly affects the dynamics of the accretion, leading to sporadic major accretion events and to discontinuous motion of the cusp of the accreting disk. This regime is known as MAD (magnetically arrested disk \citep{igumenshchev_three-dimensional_2003, narayan_magnetically_2003}), as~opposed to the SANE (standard and normal evolution, \citep{gammie_harm_2003, narayan_grmhd_2012}), where $\phi<\phi_{\text{max}}$, and the accretion is basically unaffected by the black hole magnetosphere. The~comparison with the polarized image made it possible to select  a subset of MAD models from the GRMHD simulation libraries compatible with~observations. 

Although SANE models now seem to be less favored compared to MAD models for M87*, they can still be  considered an interesting case for other sources and a perfect benchmark for GRMHD theoretical models. Recently, the~Code Comparison Project by \citet{porth_event_2019} showed the ability of the GRMHD community to produce robust modeling of the accretion in SANE disks: by adopting the same initial conditions and resolution, various state-of-the-art codes employing very different numerical strategies were able to obtain consistent results, demonstrating the maturity reached by GRMHD simulations. In~this project, the magnetic field inside the disk was initialized with a poloidal loop with a maximum magnetization, that is, the ratio of magnetic to rest mass energy densities, of~$\sigma_\text{max}\sim 10^{-2}$, while pressure perturbations were introduced to trigger the magnetorotational instability (MRI, \citep{balbus_powerful_1991,balbus_instability_1998}), as~it is needed to start the accretion process. The~output of the various codes was validated by cross-comparing results against a suite of selected diagnostics, which are currently considered a standard benchmark for this kind of~simulation.

However, such numerical models assume an initial magnetic field whose intensity is set to a high enough value as to properly resolve the onset of the MRI, rather than matching a physically relevant scenario. We can think of a more agnostic situation in which the initial magnetic field has a simpler morphology and a negligible magnetization; it is then amplified inside the disk by small-scale motions until it reaches the intensity and degree of turbulence necessary to initiate the accretion. Without~resorting to computationally expensive 3D global simulations, it is possible to achieve this effect by using the simple mean-field dynamo approach \citep{parker_hydromagnetic_1955,moffatt_magnetic_1978,krause_mean-field_1980,schrinner_mean-field_2007} even in 2D axisymmetric simulations. This is obtained by introducing in the resistive Ohm's law---a new term proportional to the magnetic field itself---which describes the coupling between the velocity and magnetic field unresolved turbulent fluctuations, with~a main effect being the creation of a poloidal magnetic component from a toroidal one (the so-called $\alpha$ effect). At~the same time, the~differential rotation of the accretion disk deforms the poloidal magnetic field lines, producing a toroidal component ($\Omega$ effect), so that the combination of these two effects allows for closing the dynamo cycle and amplifying the magnetic field exponentially (at least in the kinematic phase where the reaction of the plasma is neglected). 

 In the absence of explicit dynamo terms, this amplification process is expected to naturally take place due to the MRI: both local shearing box simulations \citep{brandenburg_dynamo-generated_1995, ziegler_shear_2001, davis_sustained_2010} and global models \citep{hawley_assessing_2011,hawley_testing_2013} studied the saturation of the MRI in order to characterize the $\alpha$ coefficients; further studies have also considered the effect of the anisotropic nature of the MHD turbulence \citep{gressel_characterizing_2015, dhang_characterizing_2020}. However, the~introduction of explicit dynamo terms allows for regulating the growth of selected field components in different spatial regions of the source, while MRI is more limited by the initial magnetic~configuration.

The covariant $3+1$ formalism for the mean-field dynamo within the context of GRMHD, together with the required specific numerical strategies, was presented by \citet{bucciantini_fully_2013} and \citet{del_zanna_covariant_2018}. The~first applications to compact objects are due to \citet{bugli_mean_2014} and \citet{franceschetti_general_2020}, both in the kinematic regime. The~first fully nonlinear study applied to accretion disks around black holes is the work by \citet{tomei_general_2020} (TO20 from here on), where the magnetic field is amplified from negligible magnetizations up to basically reaching equipartition with the plasma, hence sufficient to explain observations (synchrotron emission from Sgr A*). In~the first phase, there is the exponential growth of the magnetic field, as~already observed by \citet{bugli_mean_2014}. Later the dynamo process naturally quenches, interestingly to a level which is basically independent of the (unknown) dynamo and resistivity~coefficients.  

The main results of TO20 were recently confirmed by \citet{vourellis_relativistic_2021}, including the case of a thin accretion disk, generalizing the results of \citet{vourellis_gr-mhd_2019}. In~addition, the~mean-field dynamo approach has also been adopted  in jet launching simulations~\citep{mattia_magnetohydrodynamic_2020,mattia_magnetohydrodynamic_2020-1}, while the effects of magnetic reconnection have been studied by \mbox{\citet{ripperda_magnetic_2020}.}

The goal of the present paper is to extend the analysis of TO20 to a wider range of parameters and to~much longer dynamical times and,~above all, to reproduce the relevant standard diagnostics of the Code Comparison Project, and~hence also the EHT observations, by~using axisymmetric GRMHD simulations of accretion disks with mean-field dynamo action. The~main purpose is to show how we can correctly model the SANE accretion scenario, even adopting much cheaper 2D simulations, rather than 3D, and~this can be achieved by starting from a more agnostic, though~very unfavorable, initial configuration of a very weak (toroidal) magnetic field ($\sigma_\text{max}\sim 10^{-6}$), which is to be naturally amplified in both the toroidal and poloidal components by the action of the~dynamo.

%%%%%%%%%%%%%%%%%%%%%%%%%%%%%%%%%%%%%%%%%%
\section{Disk Model and Numerical~Setup}

In our simulations, we considered the hydrodynamic torus described by the Fishbone--Moncrief's solution
\citep{fishbone_relativistic_1976} commonly adopted in the literature, as~well as in \citet{porth_event_2019}. The~space-time is determined by a Kerr black hole with dimensionless spin $a=0.9375$ (from here on, we use units with $G=c=1$). The~inner edge of the disk is localized at $r_{in}=6M$, and the density has a maximum of $\rho = \rho_c$ at $r_c=12M$. We assumed an ideal gas equation of state with adiabatic index $\hat{\gamma}=4/3$.
Unlike TO20, here, we chose to superimpose on the hydrodynamic torus an initial toroidal magnetic field with negligible amplitude, so that the equilibrium is essentially preserved. Such a configuration is the most unfavorable to reproduce the conditions observed in ideal GRMHD models, as~they generally assume large-scale poloidal magnetic fields. We take an initial magnetic field intensity $B_\phi\propto\rho$ as in \citet{vourellis_relativistic_2021}, with~a maximum magnetization at the center of the torus of $\sigma_\text{max}=10^{-6}$. The~resistivity and dynamo profiles are similar to those in TO20, that is:
\begin{equation}
    \eta=
        \begin{cases}
            \eta_0\sqrt{\rho/\rho_c}& \rho\geq\rho_{\text{disk}}\\
            0& \rho<\rho_{\text{disk}},
        \end{cases}
\end{equation}
\begin{equation}
    \xi=
        \begin{cases}
            \xi_0 (\rho/\rho_c)\cos{\theta}& \rho\geq\rho_{\text{disk}}\\
            0& \rho<\rho_{\text{disk}},
        \end{cases}
\end{equation}
where $\eta_0$ and $\xi_0$ are the maximum values of the  resistivity and dynamo coefficients at $t=0$ (note that the density varies both in space and time),  and $\rho_{\text{disk}}=10^{-3} \rho_c$ is the density threshold that defines the boundary between disk and atmosphere. 
We also define the dynamo numbers as in \citet{vourellis_relativistic_2021}:
\begin{equation}
    C_\Omega=r\biggl{|}\frac{\partial\Omega}{\partial r}\biggr{|}\frac{H^2}{\eta},
\end{equation}
\begin{equation}
    C_\xi=H\frac{\xi}{\eta},
\end{equation}
where $\Omega=u^\phi/u^t$, and $H\simeq 0.25 r$ is the scale height of the~disk.

Using the horizon-penetrating Kerr--Schild coordinates, we carried out a series of axisymmetric simulations, expanding the range of parameters with respect to TO20 (see Table~\ref{tab1}). In~order to assess the validity of our results with respect to the diagnostics in the Code Comparison Project, we adopted the same resolutions as the simulations of \citet{porth_event_2019} (see Table~\ref{tab2} for details). Unlike in TO20, here, we did not adopt any explicit dynamo quenching. All the simulations were carried out using the ECHO code~\citep{del_zanna_echo:_2007}, with~its most recent improvements described in \citet{bugli_echo-3dhpc:_2018} and \citet{bugli_echo-3dhpc:_2018-1}.

\begin{specialtable}[H] 
\caption{Maximum initial values of $\eta$, $\xi$, $C_\Omega$, and $C_\xi$ for each~model.\label{tab1}}
%%% \tablesize{} %% You can specify the fontsize here, e.g.,~\tablesize{\footnotesize}. If commented out \small will be used.
\setlength{\tabcolsep}{8.75mm}
\begin{tabular}{ccccc}
\toprule
\textbf{Run}	& \boldmath{\textbf{$\eta_0$}}	& \boldmath{$\xi_0$} &\boldmath{$C_{\Omega,0}^{{max}}$} &\boldmath{$C_{\xi,0}^{{max}}$}\\
\midrule
Eta-3Xi-3      & $10^{-3}$           & $10^{-3}$ & $8.2\times10^{4}$ %We changed cdot to multiplication sign,  please confirm. Fine, thanks
 & $0.75$\\
Eta-3Xi-2      & $10^{-3}$           & $10^{-2}$ & $8.2\times10^{4}$ & $7.5$\\
Eta-3Xi-1      & $10^{-3}$           & $10^{-1}$ & $8.2\times10^{4}$ & $75$ \\
Eta-4Xi-3		& $10^{-4}$			& $10^{-3}$ & $8.2\times10^{5}$ & $7.5$\\
Eta-4Xi-2		& $10^{-4}$			& $10^{-2}$ & $8.2\times10^{5}$& $75$\\
Eta-4Xi-1      & $10^{-4}$           & $10^{-1}$ & $8.2\times10^{5}$ & $750$\\
\bottomrule
\end{tabular}
\end{specialtable}
\unskip

\begin{specialtable}[H] 
\caption{Details on the numerical grid adopted. Here $r_h\simeq M$ is the event horizon radius.
\label{tab2}}
%%% \tablesize{} %% You can specify the fontsize here, e.g.,~\tablesize{\footnotesize}. If commented out \small will be used.
\setlength{\tabcolsep}{8.55mm}
\begin{tabular}{cccc}
\toprule
\textbf{Axis}	& \textbf{Grid Points} & \textbf{Domain} & \textbf{Stretching}\\
\midrule
$r$ & 192 & $(r_h-0.25 M, 3000 M)$ & Logarithmic \\
$\theta$ & 192 & $(0.06, \pi-0.06)$ & Uniform \\
\bottomrule
\end{tabular}
\end{specialtable}

%%%%%%%%%%%%%%%%%%%%%%%%%%%%%%%%%%%%%%%%%%
\section{Results}

\begin{figure}[H]

\includegraphics[width=10.5 cm]{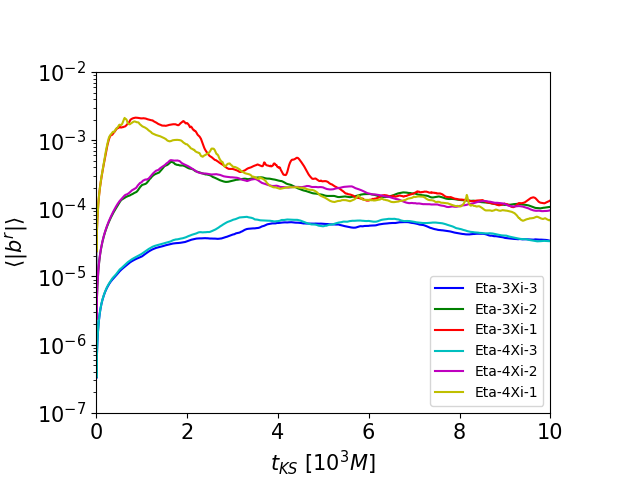}
\caption{Time evolution of $|{b^r}|$ averaged %MDPI:  Please add the comma for the number which more than five digits in figure, e.g., 16,000, 12,000, etc.
 on the disk for each~model.\label{fig_br}}
\end{figure}

We first describe the time evolution of the radial magnetic field in the torus driven by the mean-field dynamo. The~creation of a poloidal component within the disk begins from the earliest times, as~shown in Figure~\ref{fig_br}, where the temporal evolution of the disk-averaged $\langle|b^r|\rangle$, with~$|b^r|=\sqrt{b^r b^r g_{rr}}$, is plotted (we consider the comoving field). Here, the~averaging operation of any quantity $q$ is defined as in~\mbox{\citet{porth_event_2019}}:
\begin{equation}
    \langle q \rangle =\frac{ \int_{r_{\text{min}}}^{r_{\text{max}}} \int_{\theta_{\text{min}}}^{\theta_{\text{max}}}q\sqrt{-g}\,drd\theta}{{ \int_{r_{\text{min}}}^{r_{\text{max}}} \int_{\theta_{\text{min}}}^{\theta_{\text{max}}}\sqrt{-g}\,drd\theta}}
\end{equation}
regardless of $q$ being a scalar quantity or a tensorial component (as in the case we are discussing), where $r_{\text{min}}=r_h$, $r_{\text{max}}=50 M$, $\theta_{\text{min}}=\pi/3$, $\theta_{\text{max}}=2\pi/3$.

%\vspace{-12pt}
The dependence on $\xi_0$ (i.e., the dynamo $\alpha$ effect) is evident in the duration of the initial kinematic phase, in~the steepness of the curve, and~in the maximum value that is reached: the higher the $\xi_0$, the~faster the evolution towards a higher radial magnetic field value.
After a transient, there is a saturation phase followed by a decrease in $|b^r|$ inside the torus. In~particular, in~the runs of strong and intermediate dynamo, the~saturation value is similar, while it is lower in models with weak dynamo. As~shown later, the~low intensity of the initial dynamo in runs Eta-3Xi-3 and Eta-4Xi-3 seems not to be able to produce the turbulence needed to resolve the fastest growing mode of the MRI, and~thus, the accretion is entirely driven by the action of the weak~dynamo.

\begin{figure}[H]

\includegraphics[width=12 cm]{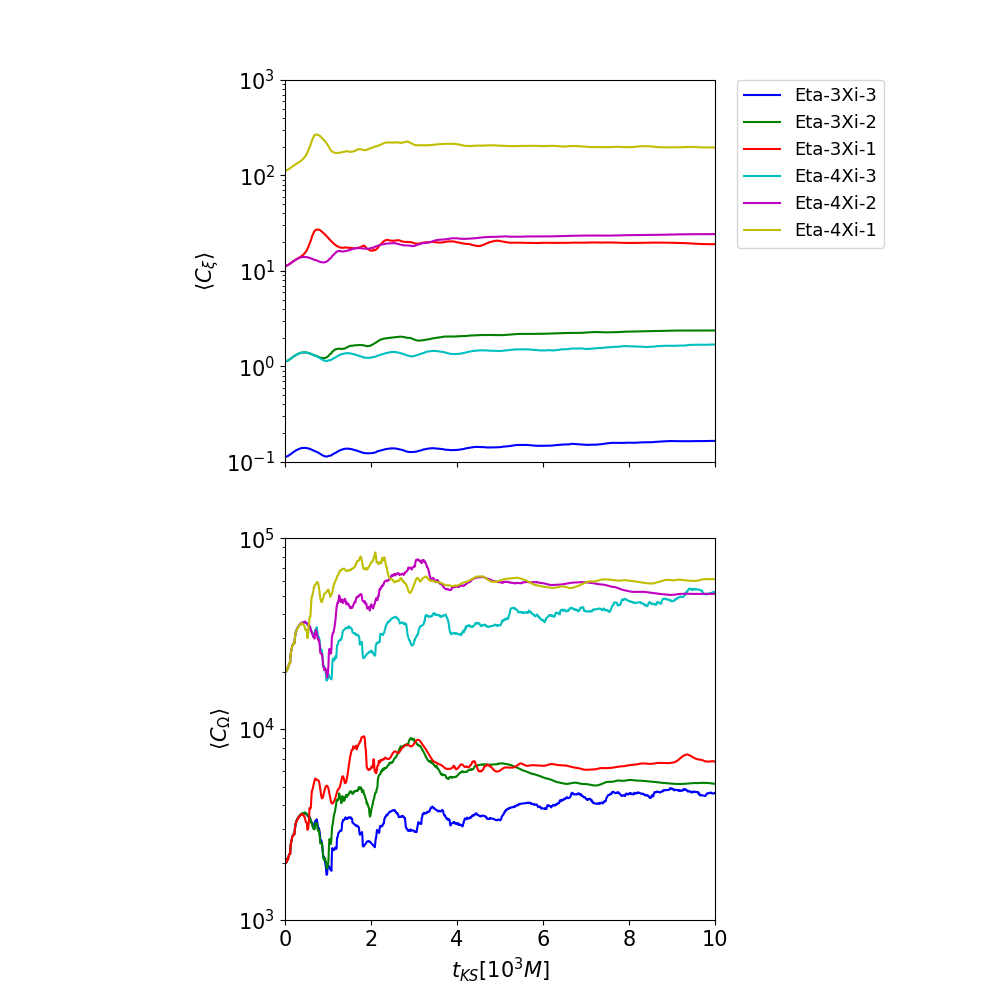}
\caption{Time evolution of the dynamo numbers averaged on the disk for each~model.\label{fig_reynolds}}
\end{figure}

The decrease of the amplification of the poloidal magnetic field in all simulations does not seem to depend on an intrinsic decrease in dynamo action.~Figure~\ref{fig_reynolds} shows that dynamo numbers remain fairly constant over time. The~most likely explanation is that the fragmentation of the magnetic structures leads to a global loss of efficiency in the dynamo coupling mechanism, hence leading naturally to a self-regulated quenching, as~first observed by TO20 and later confirmed in \citet{vourellis_relativistic_2021}.

In order to understand whether the accretion regime that is established in our dynamo runs is consistent with 3D ideal GRMHD simulations of SANE disks, in~the remainder of the paper, we attempt to reproduce the main standard diagnostics as studied in \citet{porth_event_2019}.

\subsection{Horizon Penetrating~Fluxes}
We briefly recall that the definitions of the mass, magnetic, angular momentum and energy fluxes evaluated at the horizon (see Section~5.1 in \citet{porth_event_2019}) in~the axisymmetric case are
\begin{equation}
\label{eq_mdot}
    \dot{M}= 2\pi \int_0^\pi\rho u^r \sqrt{-g} \,d\theta,
\end{equation}
\begin{equation}
    \Phi_{BH}=  \pi\int_0^\pi |^*F^{rt}|\sqrt{-g} \,d\theta,
\end{equation}
\begin{equation}
    \dot{L}=2\pi \int_0^\pi T^r_\phi \sqrt{-g} \,d\theta,
\end{equation}
\begin{equation}
\label{eq_edot}
    \dot{E}= 2\pi \int_0^\pi (-T^r_t) \sqrt{-g} \,d\theta,
\end{equation}
where $^*F^{\mu\nu}$ is the Maxwell tensor, and $T^{\mu\nu}$ is the stress-energy tensor. Usually, the fluxes defined by \eqref{eq_mdot}--\eqref{eq_edot} are normalized with the accretion rate $\dot{M}$. In~particular, the normalized magnetic flux is defined as
\begin{equation}
    \phi=\frac{\Phi_{BH}}{\sqrt{\dot{M}}}
\end{equation}
and it is also known as the MAD parameter, since we can discriminate against the SANE and MAD regimes based on its critical value. In~particular, the~critical value above which the MAD regime is favored is $\phi=15$ in the case of a black hole with spin $a = 0.9375$ and a disk with scale ratio $H/R\sim0.25$ \citep{tchekhovskoy_general_2012}.

In the ideal simulations of \citet{porth_event_2019}, the accretion starts at $t\simeq 300 \text{M}$ when the linear MRI, triggered by pressure perturbations, is captured. After~the transient, $\phi$ reaches the value $\phi\sim1$, typical of the SANE regime. Accretion rate, angular momentum flow, and net energy flow efficiency also become stationary, with~average values of $\dot{M}\sim 0.1$, $\dot{L}/\dot{M}\sim2$, and $|\dot{E}-\dot{M}|/\dot{M}\sim 0.05$.

\begin{figure}[H]
\widefigure
\includegraphics[width=16.5 cm]{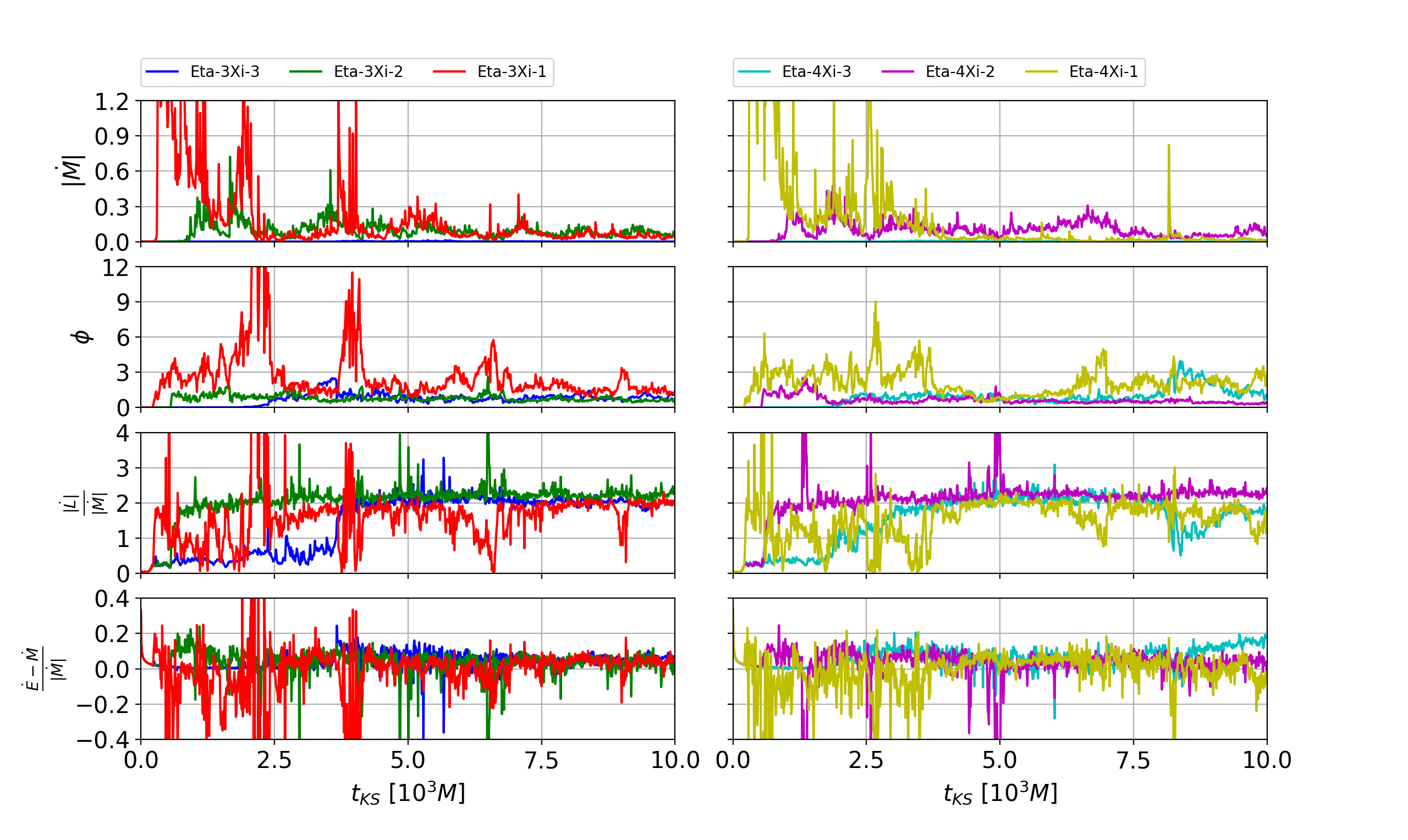}
\caption{Time series of the horizon penetrating fluxes for each model (on the left panels, runs with $\eta_0=10^{-3}$; on~the right panels, runs with $\eta_0=10^{-4}$). \label{fig_fluxes}}
\end{figure}

Figure~\ref{fig_fluxes} shows the time series of the normalized fluxes obtained in our dynamo runs. The~evolution clearly depends on the value of $\xi_0$, while there is no substantial dependence on $\eta_0$. The~two runs with weak dynamo (Eta-3Xi-3, Eta-4Xi-3) produce low accretion rates and cannot be considered good SANE models. Instead, the two runs with intermediate dynamo (Eta-3Xi-2, Eta-4Xi-2) show a very good agreement with the results shown in Figure~4 of \citet{porth_event_2019}: the accretion starts at $t\sim 1000$ M, and after~an initial transient, the~accretion rate saturates to the value of $\dot{M}\sim 0.1$. 

The magnetic flux and the angular momentum flux remain almost constant during the simulation, $\phi\sim 1$ and $\dot{L}/\dot{M}\sim 2$. The~trend in net energy flow efficiency, $(\dot{E}-\dot{M})/\dot{M}$, differs slightly from that of \citet{porth_event_2019}: in our dynamo models it can assume negative values, although~the final value is consistent with the one presented in the code comparison project, $(\dot{E}-\dot{M})/\dot{M}\lesssim 0.1$. Finally, the~two models with strong dynamo (Eta-3Xi-1, Eta-4Xi-1) show a good agreement with \citet{porth_event_2019}, although~with some differences compared to the case of intermediate dynamo. The~initial transient is characterized by a higher and more impulsive accretion~rate. 

The magnetic flux in our models with $\xi_0=0.1$ is on average higher, with~run Eta-3Xi-1 even showing two intensity spikes approaching the critical threshold of $\phi=15$. The~action of the stronger dynamo seems to affect the evolution of the angular momentum flux, which is less smooth and has values $\dot{L}/\dot{M}\lesssim2$ in the stationary phase. As~for net energy flow efficiency, an~initial phase dominated by negative values is observed, except for the two positive bursts in run Eta-3Xi-1. The~negative contribution to the energy efficiency is given by the enthalpy term \citep{mckinney_general_2012}. In~our simulations, we note that the contribution of enthalpy grows with $\xi_0$, meaning that the black hole is accreting more and more thermo-kinetically unbound~matter.

% start a new page without indent 4.6cm
\clearpage
\end{paracol}
\nointerlineskip
\vspace{-6pt}
\begin{paracol}{2}
%\linenumbers
\switchcolumn
\vspace{-6pt}

\subsection{MRI Quality~Factors}

The role of the MRI can be studied from the values assumed by the so-called quality factor. Along the $\theta$ direction, it is defined as
\begin{equation}
    Q_\theta=\frac{2\pi}{\Omega\,\sqrt{g_{\theta\theta}}\Delta\theta}\frac{|b^\theta |}{\sqrt{\rho h +b^2}},
\end{equation}
where $\rho h$ is the enthalpy per unit of volume, $\Omega$ the angular velocity of the disk, and $\Delta\theta$ the width of the cell in the $\theta$ direction.
As in TO20, we considered $\langle Q^\theta\rangle=6$ to be the threshold for resolving the MRI. Figure~\ref{fig_qtheta} shows that there is an initial time interval in which $\langle Q^\theta\rangle>6$ for runs Eta-3Xi-2, Eta-3Xi-1, Eta-4Xi-2, and Eta-4Xi-1. Then the dissipation stops the MRI and the accretion is driven by the mean-field dynamo. Run Eta-3Xi-1 has a higher value of $\langle Q^\theta\rangle$ and is kept above the threshold for a longer time than the other runs, which is consistent with the higher observed magnetic flux. Overall, according to what has been observed by TO20, it seems that MRI cannot develop for a long time with an axisymmetric mean-field dynamo mechanism: while in the initial transient, MRI and dynamo can coexist, the~subsequent stationary phase seems to be regulated exclusively by the self-quenched~dynamo.
\begin{figure}[H]

\includegraphics[width=10.5 cm]{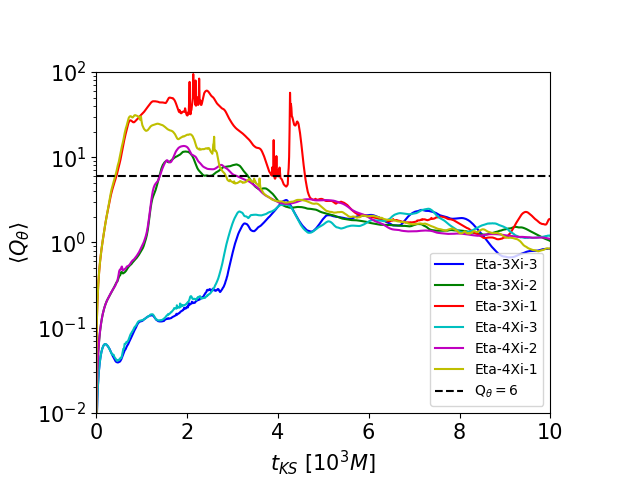}
\caption{Time evolution of the $Q_\theta$ quality factor averaged on the disk for each model. The~dashed black line indicates the minimum threshold for solving the linear~MRI.\label{fig_qtheta}}
\end{figure}   

\subsection{Time-Averaged~Maps}

To better appreciate the spatial structure of various quantities, a~picture of the quasi-stationary state can be obtained from the time averaged rest-frame density, inverse plasma $\beta$, and the magnetization $\sigma$ (Figures \ref{fig_tphi-eta-3} and \ref{fig_tphi-eta-4}). We identify the disk with the regions occupied by bound denser and weakly magnetized plasma, the~funnel by low-density and highly magnetized (Poynting-dominated) plasma, and~the jet sheath separating them and containing the larger part of the material outflows. The~time window $[5000 M,10\text{,}000 M]$ is chosen so as to cover the stationary state for all simulations, and~by adopting the same color scale used in \citet{porth_event_2019}, we will more easily compare our~results. 

The dynamo action of the dynamo significantly changes the structure of the torus.
By increasing the parameter $\xi_0$, the disk evolves to a stationary state with lower density and pressure and has a more flattened shape on the equatorial plane. The~funnel and the jet sheath are progressively wider as $\xi_0$ increases, as~the central and right panels in Figures~\ref{fig_tphi-eta-3} and \ref{fig_tphi-eta-4} show. By~increasing $C_\Omega$ by a factor of 10 (i.e., reducing the resistivity), the~contribution of the $\alpha$ effect of the dynamo during the initial transient is probably less dominant, leading the torus to lose less matter and tend more to the ideal case of \citet{porth_event_2019} (Figure \ref{fig_tphi-eta-4}).
\begin{figure}[H]

\includegraphics[width=14 cm]{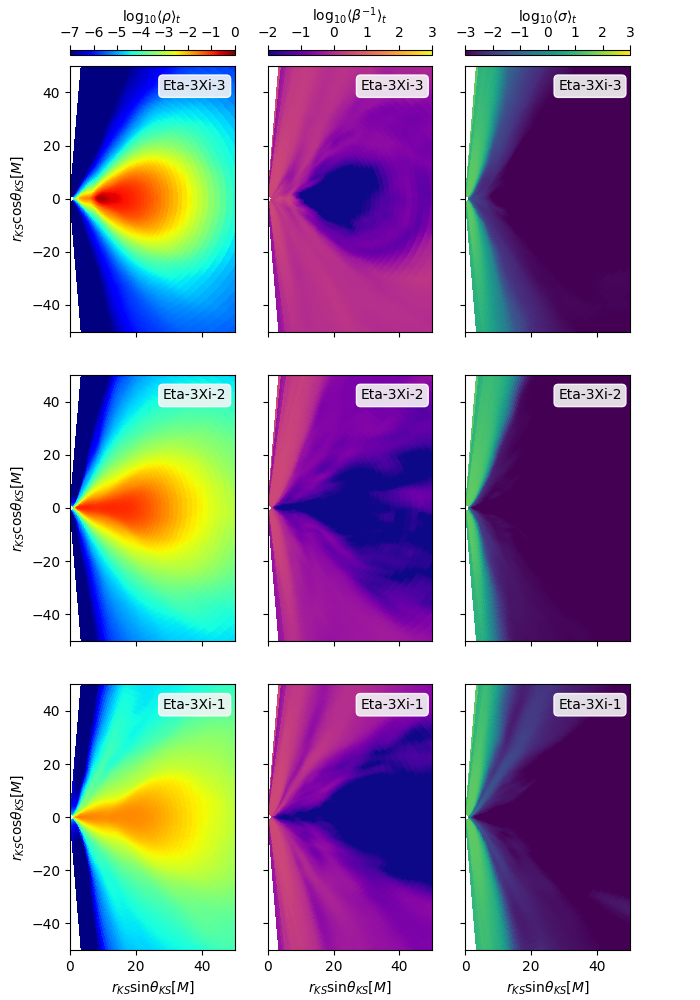}
\caption{Time-averaged data for rest-mass density, inverse plasma $\beta$, and magnetization for runs with $\eta_0=10^{-3}$. We indicate with $\langle\cdot\rangle_t$ the time average in the range [$5000 M, 10,000 M$]. %We found some ``M'' has italic format in this paper, please revise and give them a consistent choice.
\label{fig_tphi-eta-3}}
\end{figure}
\unskip   

\begin{figure}[H]

\includegraphics[width=14 cm]{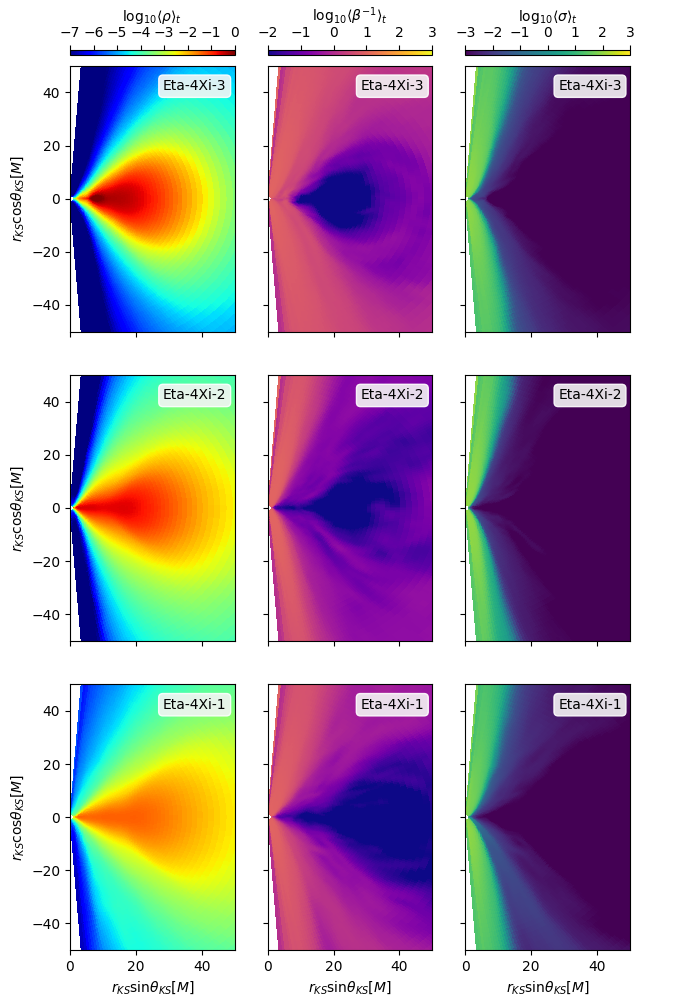}
\caption{Time-averaged data for rest-mass density, inverse plasma $\beta$, and magnetization for runs with $\eta_0=10^{-4}$.
\label{fig_tphi-eta-4}}
\end{figure}   
\subsection{Synchrotron~Pseudo-Luminosity}

A further diagnostic is a proxy for the luminosity due to (thermal) synchrotron radiation as defined in \citet{porth_event_2019}:
\begin{equation}
\mathcal{L}(t)=2\pi \int_{r_{\text{min}}}^{r_{\text{max}}} \int_{\theta_{\text{min}}}^{\theta_{\text{max}}}j(B,p,\rho)\sqrt{-g}\,d\theta dr
\end{equation}
where integration boundaries are the same as for disk-averaged quantities, and~the pseudo-emissivity is defined as $j(B,p,\rho)=\rho^3p^{-2}\exp{[-C(\rho^2/(Bp^2))^{1/3}}]$ with $C=0.2$ (all quantities in code units). Again, the~time series of the luminosity in Figure~\ref{fig_lum} shows the best agreement with \citet{porth_event_2019} in the two runs with intermediate dynamo, where $\mathcal{L}/\dot{M}\sim 200$ in the stationary phase. The~strong dynamo produces lower luminosity because the disk has lost more matter, while the weak dynamo fails to produce a magnetic field intense enough to support this~emission.

\begin{figure}[H]

\includegraphics[width=10.5 cm]{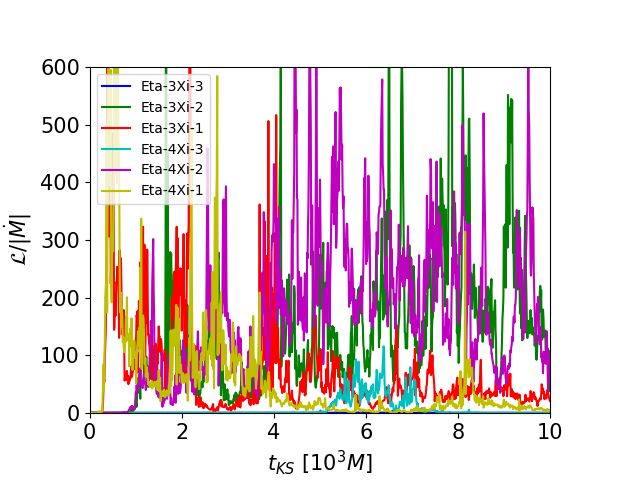}
\caption{Time evolution of the pseudo-luminosity for each model. In~runs with strong and intermediate dynamo, we see more noise than simulations of \citet{porth_event_2019}.\label{fig_lum}} 
\end{figure} 

%%%%%%%%%%%%%%%%%%%%%%%%%%%%%%%%%%%%%%%%%%
\section{Conclusions}

We have discussed here an extension of our previous work, TO20, where the first GRMHD simulations of thick accretion disks with mean-field dynamo in a dynamic regime were presented. We have reproduced the SANE disk diagnostics of the Code Comparison Project of \citet{porth_event_2019} by expanding the range of resistivity and dynamo parameters compared to our previous paper. We have also extended the duration of simulations to investigate the stationary regime in more detail. Our main purpose is to show that the approach of mean-field dynamo in axisymmetric simulations is able to reproduce the amplification of the magnetic field due to MRI-driven turbulence. Even when starting from initial conditions not particularly favorable, such as a seed (toroidal) magnetic field with very low magnetization, our models can reproduce the same results of expensive 3D GRMHD simulations after saturation and dynamical quenching of the dynamo~process. 

We can summarize the results in the following~points.
\begin{enumerate}
    \item The action of a mean-field dynamo is able to generate equipartition-like poloidal fields as already seen in TO20 and later confirmed by \citet{vourellis_relativistic_2021}.
    \item In our simulations, we find a considerable dependence on the initialization value of the $\alpha$ effect. In~particular, there is a minimum $\xi_0$ under which the dynamo is too weak to grow out the poloidal magnetic field significantly within the duration of our runs. On~the other hand, an~excessively high value causes a rapid and violent initial evolution, bringing a discrepancy of the diagnostics with the ideal three-dimensional case. However, we find an intermediate range that has a good agreement in all the diagnostics analyzed. 
    \item The dependence from $\eta_0$ is less evident in the horizon fluxes. However, the~time averages in the stationary phase show that as $\eta_0$ increases, the~average disk density decreases and the magnetization in the funnel increases. This is probably due to the fact that the action of $\alpha \Omega$ dynamo is stronger, and it more effectively influences  the accretion dynamics.
    \item We evaluated the time evolution of the $\theta$ quality factor to test the ability to solve MRI in our simulations. In~runs with intermediate and strong dynamo, the~action of the dynamo combined with the adopted resolution allows for resolving the characteristic MRI wavelength during the initial phase. If~the dynamo is too low, the~instability can never develop. This behavior is confirmed by the fact that as $\xi_0$ increases, the~accretion during the transient is progressively more intense. On~the other hand, the~stationary regime seems to be regulated exclusively by the action of the mean-field dynamo, as~also stated by \citet{vourellis_relativistic_2021}.
\end{enumerate}

In conclusion, we believe that the much more computationally convenient approach of mean-field dynamo in axisymmetric simulations can reproduce SANE accretion disks whose properties are in accordance with the fully three-dimensional ideal simulations present in the literature.
This result opens up the possibility of performing large parameter studies of axisymmetric thick disks to fit observational data and better constrain the physical properties of the accretion~flows.

For the future, we aim to investigate the behavior of SANE disks for different profiles for the resistivity and mean-field dynamo coefficients \citep{gressel_characterizing_2015} and to perform 3D simulations, so as to check whether our current results are robust, given that significant non-axisymmetric ideal and dynamo instabilities could arise \citep{bugli_papaloizou-pringle_2018}. Finally, we will extend our analysis to MAD disks, which will be the subject of the second part of the Code Comparison Project (Olivares~et~al. in prep).

\vspace{6pt}

\authorcontributions{The authors contributed equally to this work. All authors have read and agreed to the published version of the manuscript.}
\funding{MB acknowledges support from the European Research Council (ERC starting grant no. 715368---MagBURST).}%Newly added information, I moved one sentence here from the other section

\acknowledgments{The authors thank the three anonymous referees for their help in improving the manuscript. The~authors gratefully acknowledge the Gauss Centre for Supercomputing e.V. (\url{www.gauss-centre.eu}) %MDPI: Please add accessed date.
 for funding this project by providing computing time on the GCS Supercomputer SuperMUC-NG at Leibniz Supercomputing Centre (\url{www.lrz.de}).}
\conflictsofinterest{The authors declare no conflict of~interest.}

%%%%%%%%%%%%%%%%%%%%%%%%%%%%%%%%%%%%%%%%%%
\end{paracol}
\reftitle{References}

\end{document}